\documentclass[twocolumn,showpacs,preprintnumbers,amsmath,amssymb]{revtex4}
\usepackage{dsfont}
\usepackage{cases}
\usepackage{graphicx}
\usepackage{dcolumn}
\usepackage{bm}
\usepackage{amsmath}

\begin{document}

\title{Modulating quantum Fisher information of qubit in dissipative cavity by coupling strength}
\author{Danping Lin}%
\author{Yu Liu}
\author{Hong-Mei Zou}
\email{zhmzc1997@hunnu.edu.cn}%
 \affiliation{Synergetic Innovation Center for Quantum Effects and Application, Key Laboratory of Low-dimensional Quantum Structures and Quantum Control of Ministry of Education, School of Physics and Electronics, Hunan Normal University, Changsha, 410081, P.R. China.}%

\date{\today}

\begin{abstract}
By using the non-Markovian master equation, we investigate the effect of the cavity and the environment on the quantum Fisher information (QFI) of an atom qubit system in a dissipation cavity. We obtain the formulae of QFI for two different initial states and analyze the effect of the atom-cavity coupling and the cavity-reservoir coupling on the QFI. The results show that the dynamic behavior of the QFI is obviously dependent on the initial atomic states, the atom-cavity coupling and the cavity-reservoir coupling. The stronger the atom-cavity coupling, the quicker the QFI oscillates and the slower the QFI reduces. Especially, the QFI will tend to a stable value not zero if the atom-cavity coupling is large enough. On the other hand, the smaller the cavity-reservoir coupling, the stronger the non-Markovian effect, the slower the QFI decay. In other words, choosing the best parameter can improve the accuracy of parameter estimation. In addition, the physical explanation of the dynamic behavior of the QFI is given by means of the QFI flow.
\end{abstract}
\maketitle

\section{Introduction}
Quantum Fisher information (QFI), which was originally introduced by Fisher in 1925, plays an important role in quantum estimation theory and quantum information theory, and has been widely studied \cite{Fisher}. The QFI is used to describe the probability distribution of a threshold value of the parameter estimation \cite{C}. The parameter estimation in probability distribution is a very basic and essential content in information theory. Since the quantum measurement is found to be probabilistic, parameter estimation in probability distribution is gradually applied to the quantum field.  The quantum Cram\'{e}r-Rao (QCR) theorem shows that the limit of the accuracy of the parameter estimation is determined by the reciprocal of QFI \cite{C,A}. Therefore, how to increase the QFI has now become a key issue for research. In many definitions of quantum Fisher information, which is defined based on symmetric logarithmic derivative operator and is the largest of all possible classical Fisher information, i.e. this QFI takes advantage of all possible information in the state. In this paper, we will use the symmetric logarithmic derivative (SLD) method to calculate the QFI \cite{C,A}.

Recently, more attention has been paid to the QFI \cite{Zhong,Tan,Ren1,Wang,Ren2}, and QFI also has been applied widely in other quantum information tasks, such as entanglement detection \cite{Safoura}, non-Markovian description and determination\cite{Song,Ban}, and investigation of the uncertainty relations \cite{Luo,Watanabe,Sun}. The researches on the QFI mentioned are focused on the qubit in a bosonic environment\cite{Zhong}. However, in this work, we study the QFI of an atom in a dissipation cavity by using the dressed-state and the arbitrary state, and its physical significance is given. It is found that the QFI has different tendency when the strength of the coupling is different. The QFI monotonously decreases and eventually disappears in the case that the atom-cavity is weakly coupled. When the atom is strongly coupled with the cavity, the QFI will repeatedly oscillate. Furthermore, we also analyze of the dynamic behavior of QFI by means of using the QFI flow \cite{Zhong,X}.

The rest of this paper is organized as follows. In Section 2, we give a model of an atom in a dissipation cavity and the reduced density matrix of the atom. In Section 3, we introduce the QFI and the QFI flow. In Section 4, we discuss the influence of the cavity and the environment on the QFI. Finally, we end with a brief summary of important results in Section 5.

\section{Physical Model}
In this paper, we consider an atom qubit interacting with a cavity, and the cavity is coupled to a bosonic environment \cite{Zou}. The total Hamiltonian is given by
\begin{eqnarray}\label{EB01}
H&=&\frac{1}{2}\omega_{0}\sigma_{z}+\omega_{0}a^{\dag}a+\Omega(a\sigma_{+}+a^{\dag}\sigma_{-})\nonumber\\
&+&\sum_{k}\omega_{k}c_{k}^{\dag}c_{k}+(a^{\dag}+a)\sum_{k}g_{k}(c_{k}^{\dag}+c_{k})
\end{eqnarray}

where $a^{\dag}$ ($a$) is the creation (annihilation) operator of the cavity, $\sigma_{i}$ ($i=\pm, z$) is the atomic operators \cite{Jaynes}, $\omega_{0}$ is the atomic Bohr frequency and $\Omega$ is the atom-cavity coupling constant, and $c_{k}^{\dag}$ ($c_{k}$ ) is the creation (annihilation) operator of the reservoir, $g_{k}$ is the cavity-reservoir coupling strength.

Using the second order of the time convolutionless (TCL) expansion \cite{Breuer}, which neglecting the atomic spontaneous emission and the Lamb shifts, and assuming one initial excitation and a reservoir at zero temperature, the non-Markovian master equation for the density operator $R(t)$ in the dressed-state basis $\{|E_{1+}\rangle, |E_{1-}\rangle, |E_{0}\rangle\}$ is
\begin{eqnarray}\label{EB05}
\dot{R}(t)&=&-i[H_{ac},R(t)]\nonumber\\
&+&\gamma(\omega_{0}+\Omega,t)(\frac{1}{2}|E_{0}\rangle\langle E_{1+}|R(t)|E_{1+}\rangle\langle E_{0}|\nonumber\\
&-&\frac{1}{4}\{|E_{1+}\rangle\langle E_{1+}|,R(t)\})\nonumber\\
&+&\gamma(\omega_{0}-\Omega,t)(\frac{1}{2}|E_{0}\rangle\langle E_{1-}|R(t)|E_{1-}\rangle\langle E_{0}|\nonumber\\
&-&\frac{1}{4}\{|E_{1-}\rangle\langle E_{1-}|,R(t)\}),
\end{eqnarray}
where  $H_{ac}=\frac{1}{2}\omega_{0}\sigma_{z}+\omega_{0}a^{\dag}a +\Omega(a\sigma_{+}+a^{\dag}\sigma_{-})$,  $|E_{1\pm}\rangle=(|1g\rangle\pm|0e\rangle)/\sqrt{2}$ are the eigenstate of $H_{ac}$ with one total excitation, with energy $\omega_{0}/2\pm\Omega$, and $|E_{0}\rangle=|0g\rangle$ is the ground state, with energy $-\omega_{0}/2$. The timedependent decay rates for $|E_{1-}\rangle$ and $|E_{1+}\rangle$ are $\gamma(\omega_{0}-\Omega,t)$ and $\gamma(\omega_{0}+\Omega,t)$ respectively.

If the reservoir at zero temperature is modeled with a Lorentzian  spectral density\cite{Scala1,Zou1}
\begin{equation}\label{EB06}
J(\omega)=\frac{1}{2\pi}\frac{\gamma_{0}\lambda^{2}}{(\omega_{1}-\omega)^{2}+\lambda^{2}},
\end{equation}
where the parameter $\lambda$ defines the spectral width of the coupling, which is connected to the reservoir correlation time $\tau_{R}$ by $\tau_{R}$=$\lambda^{-1}$ and the parameter $\gamma_{0}$ is related to the relaxation time scale $\tau_{S}$ by $\tau_{S}$=$\gamma_{0}^{-1}$. $\lambda>2\gamma_{0}$ is called a weak coupling regime or a Markvian regime. In this regime, the relaxation time is greater than the reservoir correlation time. Supposing the spectrum is peaked on the frequencies of the states $|E_{1-}\rangle$, i.e. $\omega_{1}=\omega_{0}-\Omega$, the decay rates for the two dressed states $|E_{1\pm}\rangle$ are respectively expressed as\cite{Scala1} $\gamma(\omega_{0}-\Omega,t)=\gamma_{0}(1-e^{-\lambda t})$ and $\gamma(\omega_{0}+\Omega,t)=\frac{\gamma_{0}\lambda^{2}}{4\Omega^{2}+\lambda^{2}}\{1+[\frac{2\Omega}{\lambda}sin2\Omega t-cos2\Omega t]e^{-\lambda t}\}$.

We can acquire the density matrix elements of the atom-cavity at all times from Eq.~(\ref{EB05})
\begin{eqnarray}\label{EB10}
      R_{11}(t)&=&A_{11}^{11}R_{11}(0),  R_{12}(t)=A_{12}^{12}R_{12}(0),   \nonumber\\    R_{13}(t)&=&A_{13}^{13}R_{13}(0),\nonumber\\
      R_{22}(t)&=&A_{22}^{22}R_{22}(0),      R_{23}(t)=A_{23}^{23}R_{23}(0),\nonumber\\
      R_{33}(t)&=&A_{33}^{11}R_{11}(0)+A_{33}^{22}R_{22}(0)+A_{33}^{33}R_{33}(0),
\end{eqnarray}
where $R_{ij}(0) (i,j=1,2,3)$ is the density matrix element of the initial state, and
\begin{eqnarray}\label{EB11}
      A_{11}^{11}&=&e^{-\frac{1}{2}I_{+}},   A_{12}^{12}=e^{-2i\Omega t}e^{-\frac{1}{4}(I_{+}+I_{-})},\nonumber\\  A_{13}^{13}&=&e^{-i(\omega_{0}+\Omega)t}e^{-\frac{1}{4}I_{+}},\nonumber\\
      A_{22}^{22}&=&e^{-\frac{1}{2}I_{-}},   A_{23}^{23}=e^{-i(\omega_{0}-\Omega)t}e^{-\frac{1}{4}I_{-}},\nonumber\\
      A_{33}^{11}&=&1-A_{11}^{11},   A_{33}^{22}=1-A_{22}^{22},      A_{33}^{33}=1,
\end{eqnarray}
and
\begin{eqnarray}\label{EB12}
     I_{-}&=&\gamma_{0}t+\frac{\gamma_{0}}{\lambda}(e^{-\lambda t}-1),\nonumber\\
     I_{+}&=&\frac{\gamma_{0}\lambda^{2}}{4\Omega^{2}+\lambda^{2}}[t-\frac{4\Omega e^{-\lambda t} sin(2\Omega t)}{4\Omega^{2}+\lambda^{2}}\nonumber\\ 
     &+&\frac{(\lambda^{2}-4\Omega^{2})(e^{-\lambda t}cos(2\Omega t)-1)}{\lambda(4\Omega^{2}+\lambda^{2})}].
\end{eqnarray}

By taking a partial trace of the atom-cavity density matrix over the cavity degree of freedom, the atomic reduced operator $\rho(t)$ is given by \cite{Yu}
\begin{eqnarray}\label{EB13}
\rho(t)=\left(
            \begin{array}{cccc}
              \rho_{11}(t)&\rho_{12}(t)\\
              \rho_{21}(t)&\rho_{22}(t)\\

            \end{array}
          \right),
\end{eqnarray}
here
\begin{eqnarray}\label{EB14}
      \rho_{11}(t)&=&\frac{1}{2}(R_{11}(t)+R_{12}(t)+R_{21}(t)+R_{22}(t)+2R_{33}(t)),\nonumber\\      \rho_{12}(t)&=&\frac{1}{\sqrt{2}}(R_{13}(t)-R_{23}(t)),\nonumber\\
      \rho_{21}(t)&=&\frac{1}{\sqrt{2}}(R_{31}(t)-R_{32}(t)),\nonumber\\      \rho_{22}(t)&=&\frac{1}{2}(R_{11}(t)-R_{12}(t)-R_{21}(t)+R_{22}(t)),\nonumber\\
\end{eqnarray}

\section{Quantum Fisher Information}

\subsection{Quantum Fisher Information}
The QFI indicates the sensitivity of the state to the change of the parameter. Let $\phi$ denote a single parameter to be estimated, the QFI is defined as \cite{A}
\begin{eqnarray}\label{EB15}
     \begin{array}{cccc}
      F_{\phi}&=&Tr(\rho_{\phi}L_{\phi}^2)=Tr(\partial_{\phi}\rho_{\phi}L_{\phi})\\
     \end{array},
\end{eqnarray}
where $L_{\phi}$ is symmetric logarithmic derivative(SLD) for the parameter $\phi$, which is a Hermitian operator determined by
\begin{eqnarray}\label{EB16}
     \begin{array}{cccc}
      \partial_{\phi}\rho_{\phi}&=&\frac{1}{2}\{\rho_{\phi},L_{\phi}\}\\
     \end{array},
\end{eqnarray}
where $\partial_{\phi}\equiv\frac{\partial}{\partial\phi}$ and $ \{\cdot,\cdot\}$ denotes the anticommutator.

An essential feature of the QFI is that we can obtain the achievable lower bound of the mean-square error of unbiased estimators for the parameter $\phi$ through the quantum Cram\'{e}r-Rao (QCR) theorem \cite{Pairs}
\begin{eqnarray}\label{EB17}
     \begin{array}{cccc}
     V_{ar}(\phi)&\geq&\frac{1}{\nu F_{\phi}}\\
     \end{array},
\end{eqnarray}
where $V_{ar}(\cdot)$ denotes the variance, $\phi$ denotes the unbiased estimator, and $\nu$ denotes the number of repeated experiments. In the following, we use this method to calculate the QFI of the atom coupled to the dissipation cavity.

In order to understand the dynamic behavior of QFI, we introduce the QFI flow, which is defined as the change rate of the QFI by \cite{X}
\begin{eqnarray}\label{EB18}
     \begin{array}{cccc}
      I_{\phi}&=&\frac{\partial  F_{\phi}}{\partial t}\\
     \end{array},
\end{eqnarray}
It is well known that $I_{\phi}<0$ represents the energy and information flow from the system to the environment, and $I_{\phi}>0$ represents the energy and information flow from the environment to the system.

\subsection{Example}
Example 1. We construct a complete orthogonal basis for a three-state system with the dressed-state basis $ \{|E_{1+}\rangle, |E_{1-}\rangle, |E_0\rangle\}$,
\begin{eqnarray}\label{EB19}
     \begin{array}{cccc}
      |\psi_{1}\rangle&=&\frac{1}{\sqrt{2}}e^{i\phi} sin\frac{\theta}{2}|E_{1+}\rangle-\frac{1}{\sqrt{2}}e^{i\phi} sin\frac{\theta}{2}|E_{1-}\rangle+cos\frac{\theta}{2}|E_0\rangle\\
     \end{array},
\end{eqnarray}
where $\theta \in[0,\pi)$ and $\phi\in[0,2\pi)$ are parameters which may be regarded as encoding the amplitude and phase information, respectively.

Inserting  Eq.~(\ref{EB19}) into Eq.~(\ref{EB13}), the reduced density matrix of the atom is given by
\begin{eqnarray}\label{EB20}
\rho'(t)=\left(
            \begin{array}{cccc}
              \rho'_{11}(t)&\rho'_{12}(t)\\
              \rho'_{21}(t)&\rho'_{22}(t)\\

            \end{array}
          \right),
\end{eqnarray}
where the matrix elements are
\begin{eqnarray}\label{EB21}
      \rho'_{11}(t)&=&[1-\frac{1}{4}(A_{11}^{11}-A_{12}^{12}-A_{21}^{21}+A_{22}^{22})]sin^2\frac{\theta}{2}+cos^2\frac{\theta}{2},\nonumber\\      \rho'_{12}(t)&=&\frac{1}{2}(A_{13}^{13}+A_{23}^{23})e^{-i\phi} sin\frac{\theta}{2}cos\frac{\theta}{2},\nonumber\\
      \rho'_{21}(t)&=&\frac{1}{2}(A_{31}^{31}+A_{32}^{32})e^{i\phi} sin\frac{\theta}{2}cos\frac{\theta}{2},\nonumber\\      \rho'_{22}(t)&=&\frac{1}{4}(A_{11}^{11}-A_{12}^{12}-A_{21}^{21}+A_{22}^{22})sin^2\frac{\theta}{2},\nonumber\\
\end{eqnarray}
From Eq.~(\ref{EB15}) and Eq.~(\ref{EB20}), we obtain the QFI of the parameter $\phi$
\begin{eqnarray}\label{EB22}
     \begin{array}{cccc}
      F_{\phi}&=&\frac{1}{4}(A_{13}^{13}+A_{23}^{23})(A_{31}^{31}+A_{32}^{32})sin^{2}\theta\\
     \end{array},
\end{eqnarray}
Example 2. We consider a simple example to calculate the QFI. We use the standard basis $|e\rangle\equiv(1,0)^{T}$ and $|g\rangle\equiv(0,1)^{T}$, corresponding to the ground state and excited state, respectively. Consider an arbitrary single-qubit state
\begin{eqnarray}\label{EB23}
     \begin{array}{cccc}
      |\psi_{2}\rangle&=&cos\frac{\theta}{2}|e\rangle+e^{i\phi} sin\frac{\theta}{2}|g\rangle\\
     \end{array},
\end{eqnarray}
where $\theta$ and $\phi$ refer to the polar and azimuth angles on the Bloch sphere. Here, two parameters $\theta$ and $\phi$ in  Eq.~(\ref{EB23}) are assumed to unitary encoded.

With Eq.~(\ref{EB13}) and Eq.~(\ref{EB23}), the atom reduced density matrix $\rho(t)$ is as
\begin{eqnarray}\label{EB24}
\rho''(t)=\left(
            \begin{array}{cccc}
              \rho''_{11}(t)&\rho''_{12}(t)\\
              \rho''_{21}(t)&\rho''_{22}(t)\\

            \end{array}
          \right),
\end{eqnarray}
and the matrix elements are
\begin{eqnarray}\label{EB25}
      \rho''_{11}(t)&=&[1-\frac{1}{4}(A_{11}^{11}-A_{12}^{12}-A_{21}^{21}+A_{22}^{22})]cos^2\frac{\theta}{2}+sin^2\frac{\theta}{2},\nonumber\\      \rho''_{12}(t)&=&\frac{1}{2}(A_{13}^{13}-A_{23}^{23})e^{-i\phi} sin\frac{\theta}{2}cos\frac{\theta}{2},\nonumber\\
      \rho''_{21}(t)&=&\frac{1}{2}(A_{31}^{31}-A_{32}^{32})e^{i\phi} sin\frac{\theta}{2}cos\frac{\theta}{2},\nonumber\\      \rho''_{22}(t)&=&\frac{1}{4}(A_{11}^{11}-A_{12}^{12}-A_{21}^{21}+A_{22}^{22})cos^2\frac{\theta}{2},\nonumber\\
\end{eqnarray}
Using Eq.~(\ref{EB15}) and Eq.~(\ref{EB23}), we obtain the QFI of the parameter $\phi$
\begin{eqnarray}\label{EB26}
     \begin{array}{cccc}
      F'_{\phi}&=&\frac{1}{4}(A_{13}^{13}-A_{23}^{23})(A_{31}^{31}-A_{32}^{32})sin^{2}\theta\\
     \end{array},
\end{eqnarray}

\section{Discussion and results}
Now we use Eq.~(\ref{EB22}) and Eq.~(\ref{EB26}) to calculate the QFI of the parameter $\phi$. Because the atom is coupled to the dissipation cavity, both of the atom-cavity coupling constant and the cavity-reservoir coupling strength can affect on the dynamic behavior of the QFI on the parameter $\phi$. We study the effect of the cavity-reservoir coupling strength on the QFI under both Markovian and non-Markovian regimes. In the meantime, we also study the effect of the atom-cavity coupling constant on the QFI under weak and strong coupling.

Let us begin with the first example by taking the dressed-state $|\psi_{1}\rangle=\frac{1}{\sqrt{2}}e^{i\phi} sin\frac{\theta}{2}|E_{1+}\rangle-\frac{1}{\sqrt{2}}e^{i\phi} sin\frac{\theta}{2}|E_{1-}\rangle+cos\frac{\theta}{2}|E_0\rangle$ of the atom coupled to the dissipation cavity, we first concentrate on $\theta=\frac{\pi}{2}$. Figure. 1 shows that the QFI dynamics of the atom coupled to dissipation cavity with the dressed-state $|\psi_{1}\rangle$ in the Markovian ($\lambda=5\gamma_{0}$) and the non-Markovian ($\lambda=0.05\gamma_{0}$) regimes.

\begin{center}
\includegraphics[width=6cm,height=3.5cm]{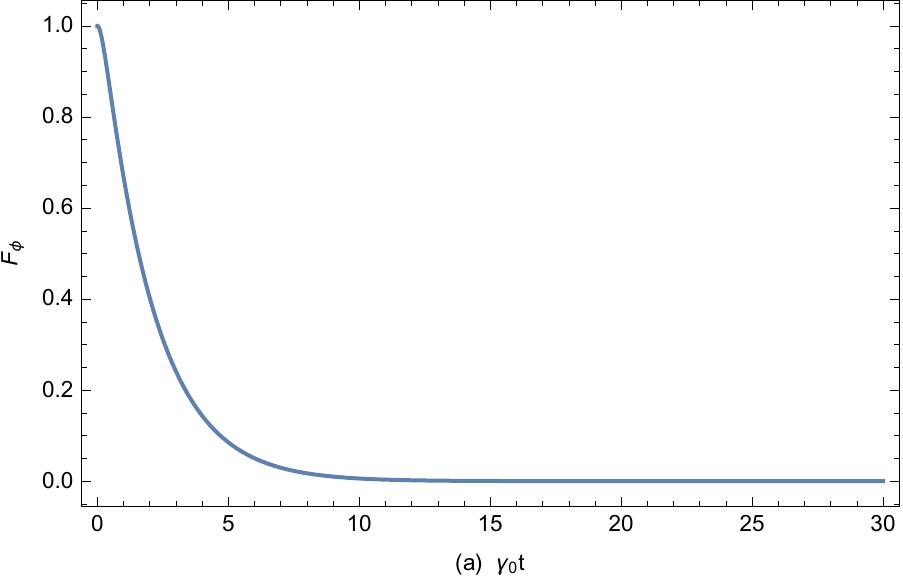}
\includegraphics[width=6cm,height=3.5cm]{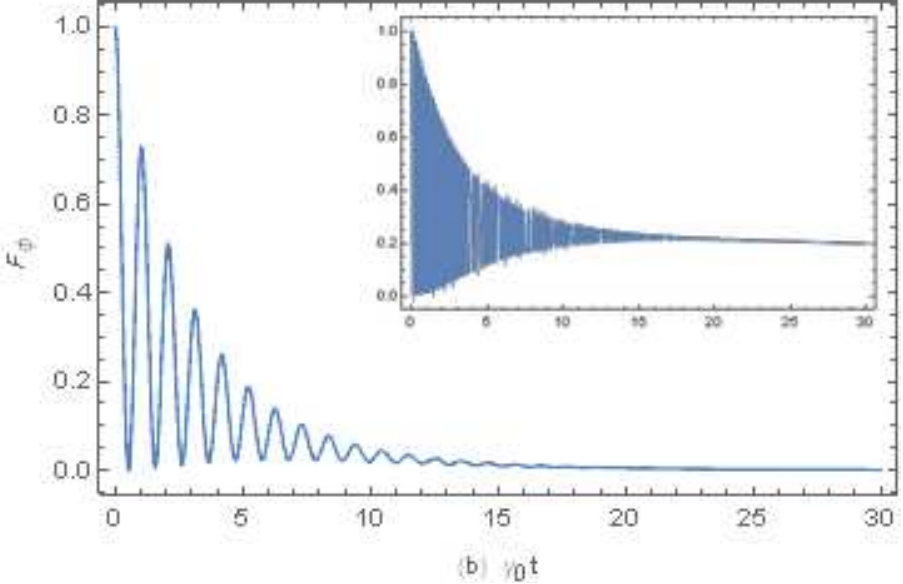}
\includegraphics[width=6cm,height=3.5cm]{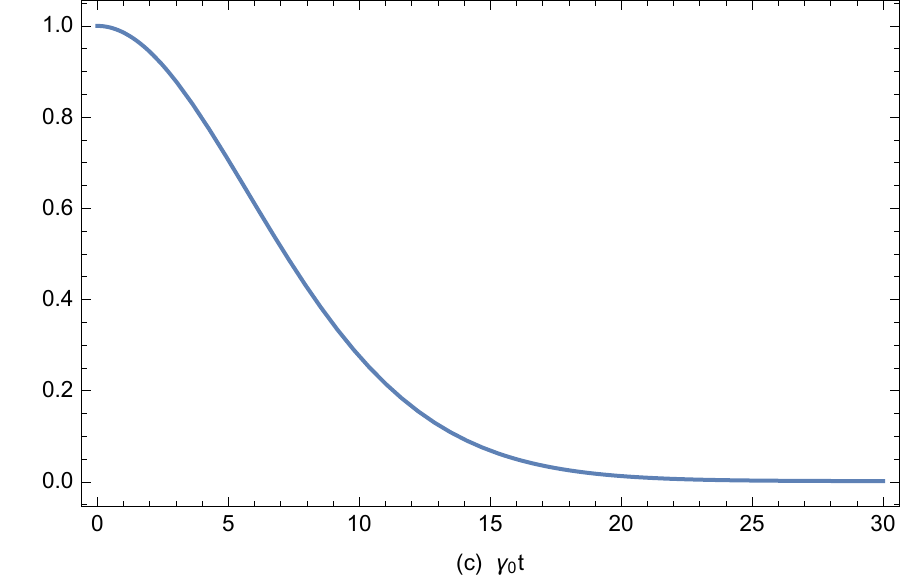}
\includegraphics[width=6cm,height=3.5cm]{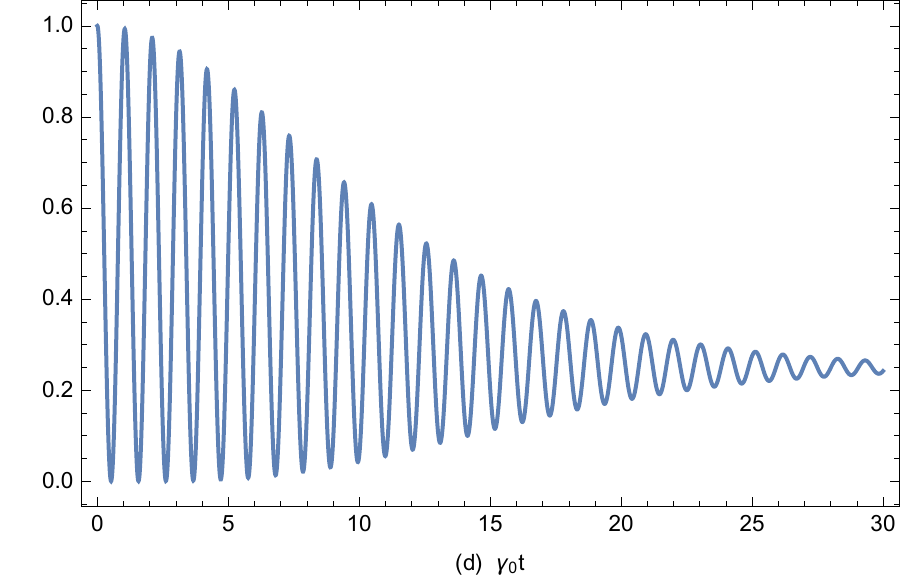}
\parbox{8cm}{\small{\bf Figure1.}
(Color online) $F_\phi$ as a function of the dimensionless quantity $\gamma_0t$ for the initial state $|\psi_{1}\rangle$. Here $\theta=\frac{\pi}{2}$. (a)$\lambda=5\gamma_{0}$, $\Omega=0.05\gamma_{0}$; (b)$\lambda=5\gamma_{0}$, $\Omega=3\gamma_{0}$; (c)$\lambda=0.05\gamma_{0}$, $\Omega=0.05\gamma_{0}$; (d)$\lambda=0.05\gamma_{0}$, $\Omega=3\gamma_{0}$. The inset in (b) shows the case of $\Omega=20\gamma_{0}$. }
\end{center}

As we know, $\lambda>2\gamma_{0}$ represents the  Markovian regime of the reservoir, $\lambda<2\gamma_{0}$ represents the non-Markovian regime of the reservoir. $\Omega>2\gamma_{0}$ represents the strong atom-cavity coupling, and $\Omega<2\gamma_{0}$ represents the weak atom-cavity coupling. As is plotted in Figure.1 (a), the reservoir is Markovian ($\lambda=5\gamma_{0}$) and the atom is a weak coupling with the cavity ($\Omega=0.05\gamma_{0}$), $F_\phi$ decreases monotonously with time and quickly close to zero. In Figure.1 (b), $\lambda=5\gamma_{0}$ (the Markovian reservoir), but $\Omega=3\gamma_{0}$, that is, the atom-cavity coupling is strong, $F_\phi$ oscillates damply with time. Then with the increase of time, $F_\phi$ again rises to 0.75 from zero. After several cycles of oscillation, $F_\phi$ ultimately decays to zero. Comparing Figure.1 (a) and Figure.1 (b), we know that the energy and information can be swapped effectively between the atom and the cavity when $\Omega=3\gamma_{0}$ and $\lambda=5\gamma_{0}$. Thus $F_\phi$ will oscillate significantly and $F_\phi$ final decay to zero due to the dissipation of the Markovian reservoir. However, $F_\phi$ will tend to a stable value when $\Omega=20\gamma_{0}$, showed the inset of Figure. 2(b). Figure.1 (c) indicates the dynamic behavior of $F_{\phi}$ in the non-Markovian regime ($\Omega=0.05\gamma_{0}$) and the weak atom-cavity coupling ($\lambda=5\gamma_{0}$). $F_\phi$ also reduces monotonously with time and vanishes only in the asymptotic limit $t\rightarrow\infty$. Comparing Figure.1 (a) and Figure.1 (c), it is seen that the decay rate of $F_\phi$ in the former is obviously larger than in the latter.  As we can see from Figure.1 (d), when $\lambda=0.05\gamma_{0}$ (in the non-Markovian regime) and $\Omega=3\gamma_{0}$ (the strong atom-cavity), $F_\phi$ will tend to a stable value and not zero after many cycles of oscillation. Comparing Figure.1 (b) and Figure.1 (d), it is found that, in the same atom-cavity coupling ($\Omega=3\gamma_{0}$), due to the memory and feedback effect of the non-Markovian reservoir, the time of oscillation in Figure.1 (d) is larger than in Figure.1 (b), and the stable value in Figure.1 (d) is bigger than in Figure.1 (b).

\begin{center}
\includegraphics[width=6cm,height=3.5cm]{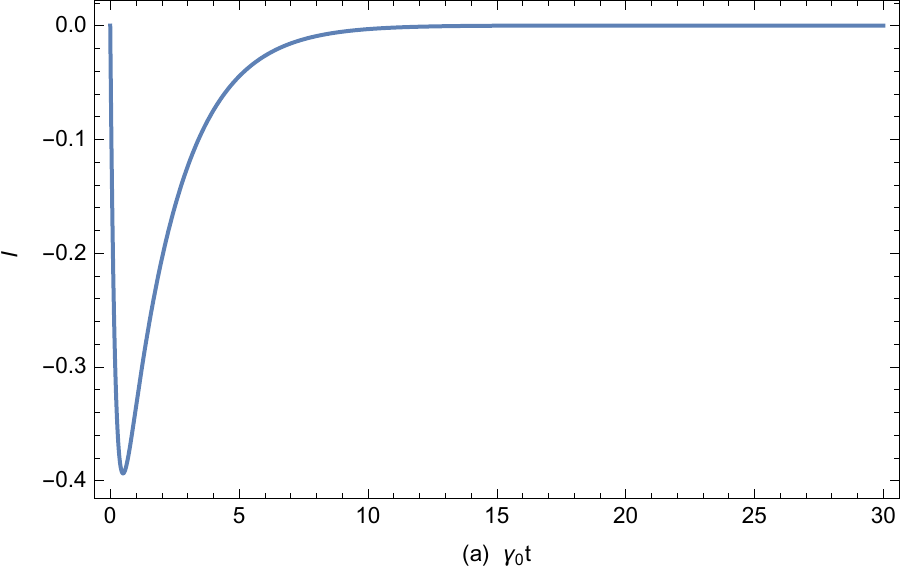}
\includegraphics[width=6cm,height=3.5cm]{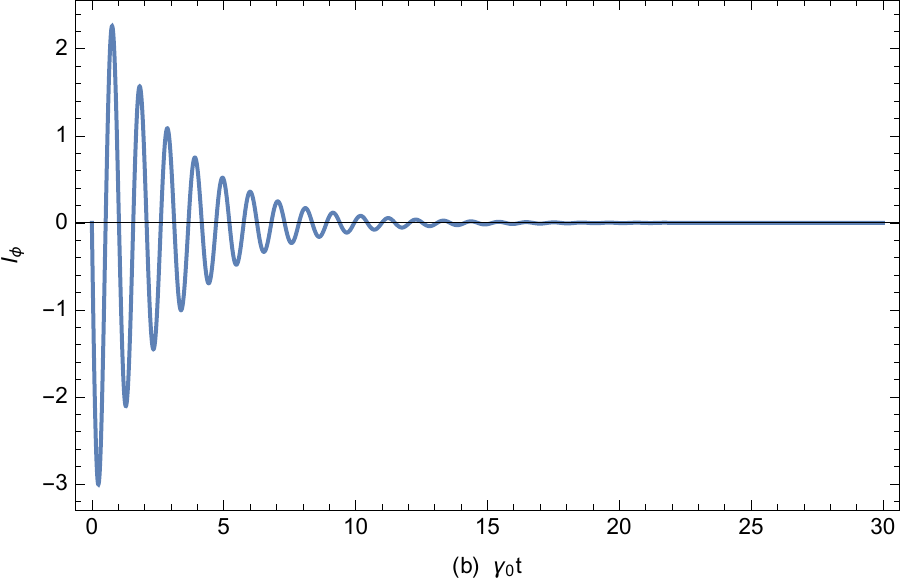}
\includegraphics[width=6cm,height=3.5cm]{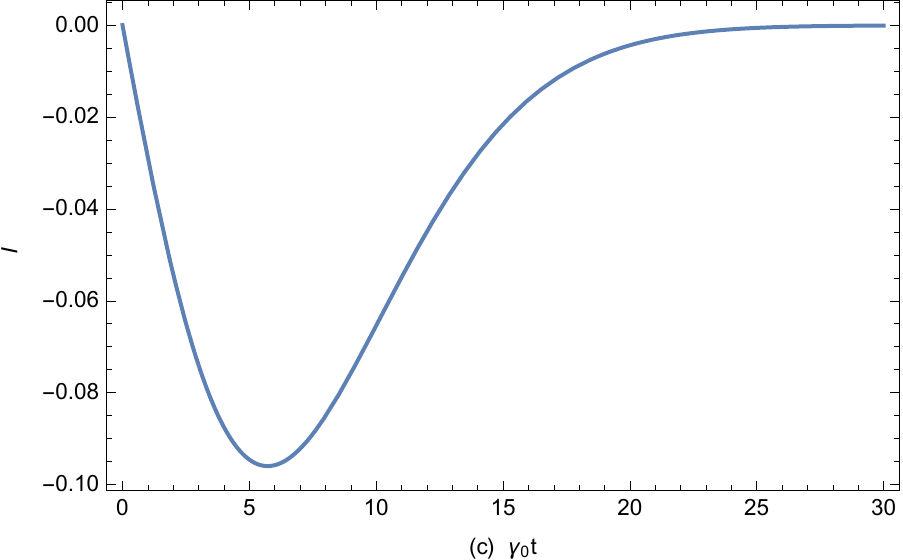}
\includegraphics[width=6cm,height=3.5cm]{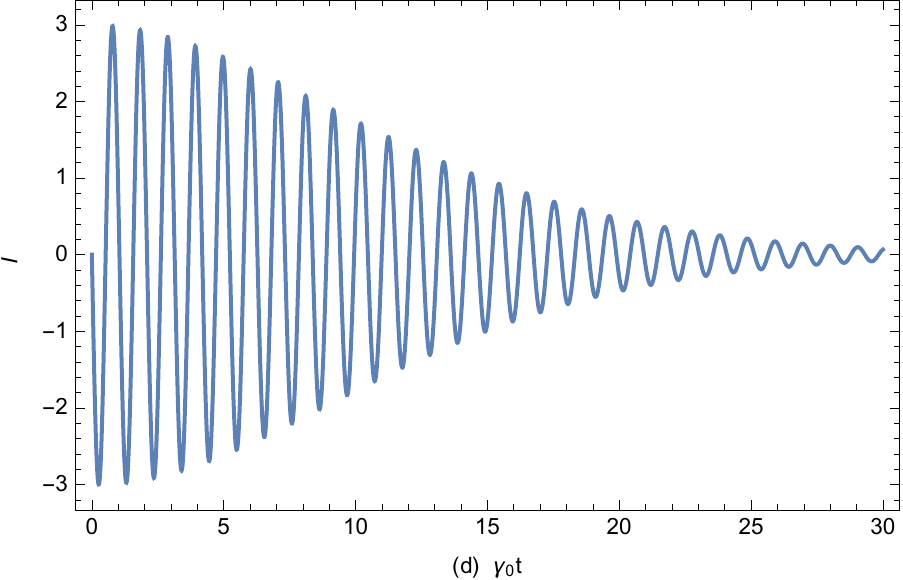}
\parbox{8cm}{\small{\bf Figure2.}
(Color online)The effect of the atom-cavity coupling $\Omega$ and the cavity-reservoir coupling $\lambda$ on the QFI flow ($I_{\phi}$)versus $\gamma_{0}t$. Here $\theta=\frac{\pi}{2}$. Here $\theta=\frac{\pi}{2}$. (a)$\lambda=5\gamma_{0}$, $\Omega=0.05\gamma_{0}$; (b)$\lambda=5\gamma_{0}$, $\Omega=3\gamma_{0}$; (c)$\lambda=0.05\gamma_{0}$, $\Omega=0.05\gamma_{0}$; (d)$\lambda=0.05\gamma_{0}$, $\Omega=3\gamma_{0}$. }
\end{center}

In Figure.2, we plot the QFI flow as a function of $\gamma_{0}$t with the initial state $|\psi_{1}\rangle$ in the Markovian and non-Markovian regime for different numbers of $\Omega$. From Figure.2 (a), we can find that $I_{\phi}$ changes rapidly to $-0.4$ from zero, then again rise to zero when $\lambda=5\gamma_{0}$ and $\Omega=0.05\gamma_{0}$. That is, $I_{\phi}$ is always less than zero. This shows that the energy and information quickly flow to the reservoir from the atom so that $F_{\phi}$ quickly reduces zero. This result is consistent with Figure.1 (a). In Figure.2 (b), $I_{\phi}$ reduces very fast to 3.0 from zero then again rises very fast to +2.3, and then becomes damping oscillation and closes to zero. Because $I_{\phi}<0$ indicates the energy and information flow out from the atom and $I_{\phi}>0$ represents the energy and information flow in the atom, $F_{\phi}$ quickly decreases when $I_{\phi}$ changes from $0\rightarrow-0.5\rightarrow0$, and $F_{\phi}$ again rise quickly to a certain value when $I_{\phi}$ changes from $0\rightarrow2.3\rightarrow0$, then $F_{\phi}$ oscillates to zero, as shown in Figure.1 (b). In addition, comparing Figure.2 (b) and (d), it can be seen that the trend of QFI flow is similar. The difference is the value of $\lambda$. In particular, the smaller the value of $\lambda$ is, the strong the non-Markovian effect is, the slower the evolution of $I_{\phi}$ decay. By comparing Figure.2 (a) and (b), Figure.2 (c) and (d), we can see that the larger the value of $\Omega$ is, the stronger the cavity-reservoir coupling is, the faster the QFI flow oscillate and recover. These phenomena can be understood as the reverse flow of energy and information from the reservoir to the system. As a result, QFI is linked to the flow of information exchanged between the system and the environment.

As the second example, we consider an arbitrary single-qubit state $ |\psi_{2}\rangle=cos\frac{\theta}{2}|e\rangle+e^{i\phi} sin\frac{\theta}{2}|g\rangle$ in the standard basis. In order to analyze the effect of the atom-cavity coupling $\Omega$ in the Markovian ($\lambda=5\gamma_{0}$) and non-Markovian ($\lambda=0.05\gamma_{0}$) regimes, we plot the evolution of $F_{\phi}$ on the $\theta=\frac{\pi}{2}$ in Figure. 3. We can observe that, in Figure. 3 (a), the reservoir is Markovian ($\lambda=5\gamma_{0}$) and the atom is a weak coupling with the cavity ($\Omega=0.05\gamma_{0}$), $F_{\phi}$ rises to 0.08 from zero then oscillates damply to zero. Beside this, this figure also shows that with the increase of $\gamma_{0}t$, the value of $F_{\phi}$ drops considerably, which implies that the accuracy of the estimate is higher. In Figure. 3(b), it can be found that, when $\lambda=5\gamma_{0}$ (the Markovian reservoir) and $\Omega=3\gamma_{0}$, that is the atom-cavity coupling is strong, $F_{\phi}$ will oscillate increases to 1.0 from zero. In particular, when $\Omega=20\gamma_{0}$, $F_{\phi}$ will oscillate damply to the stable value of 0.25 not zero, showed as the inset in Figure. 3(b). Comparing Figure. 3(a) and Figure. 3(c), we can see that their $F_{\phi}$ dynamics are similar. The difference is the maximum of $F_{\phi}$. The latter has a bigger maximum and tends to zero slower. As shown in the Figure. 3(b) and Figure. 3(d), the trend of $F_{\phi}$ is similar, except that the former eventually reaches a stable value of 0.25, and the latter disappears for the smaller $\Omega$, but when $\Omega=20\gamma_{0}$，$F_{\phi}$ will tend to 0.25. In Figure. 3, with the increase of $\Omega$, the oscillating frequency of $F_{\phi}$ will become quick and the decay of $F_{\phi}$ will become slow. But in the weak coupling, all QFIs will eventually reduce to zero in a short time, shown as Figure. 3(a) and Figure. 3(c). However, in the strong coupling regime, all $F_{\phi}$ will quickly oscillation and tend to the stable value of 0.25 in the end.

In Figure. 4, we plot the effect of the atom-cavity coupling $\Omega$ on the $I_{\phi}$ dynamics in the Markovian ($\lambda=0.05\gamma_{0}$) and non-Markovian ($\lambda=5\gamma_{0}$) regimes. In Figure. 4(a), it can be found that, when $\lambda=0.05\gamma_{0}$ (the Markovian regime), $\Omega=3\gamma_{0}$ (i.e., the weak atom-cavity coupling), as time $t$ increases, $I_{\phi}$ will increase from zero to 0.0025 and then oscillates to zero. This shows that the atom first obtains the energy and information from the cavity by the atom-cavity then this energy and information will disappear, due to the cavity dissipation. This is corresponding to the oscillation of $F_{\phi}$, seeing Figure. 3(a). In addition, it can be found that the $I_{\phi}$ quickly oscillates to zero in the strong coupling in Figure. 4(b). This is accounting for the oscillation of $F_{\phi}$, seeing Figure. 3(b). Because the cavity has memory effect, and some of the missing information can be returned to the system of the cavity. In Figure. 4(c), $I_{\phi}$ oscillates slowly to zero. Comparing Figure. 4(a) and Figure. 4(c), we can see that, the bigger the value of $\Omega$ is, the slower $I_{\phi}$ decay to zero. In Figure. 4(d), $I_{\phi}$ quickly oscillates and decays to zero, which indicates that the information of the system and environment are quickly interflowing.

\begin{center}
\includegraphics[width=6cm,height=3.5cm]{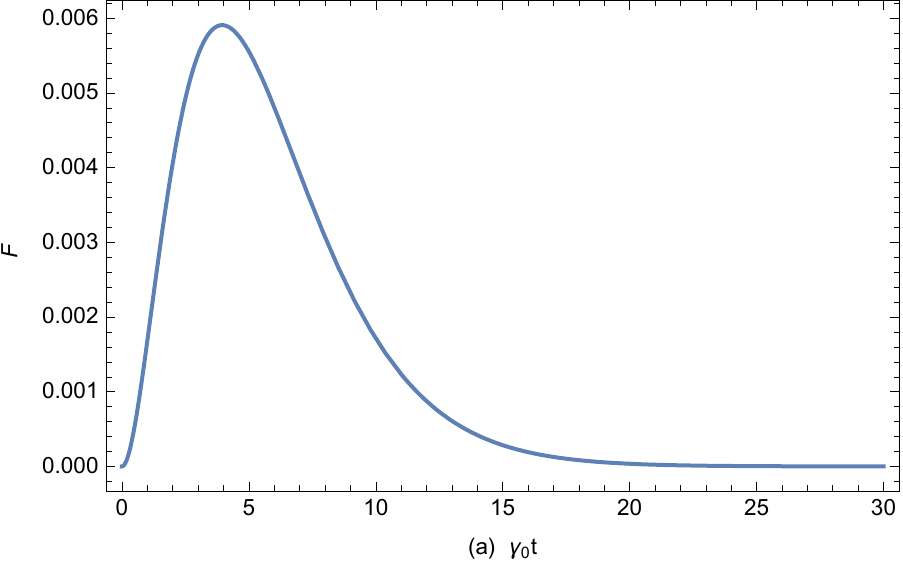}
\includegraphics[width=6cm,height=3.5cm]{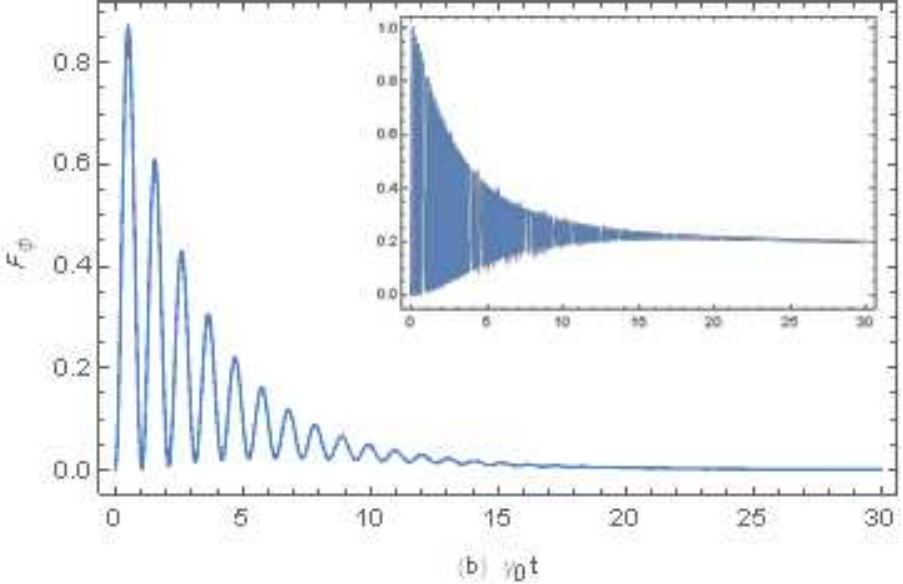}
\includegraphics[width=6cm,height=3.5cm]{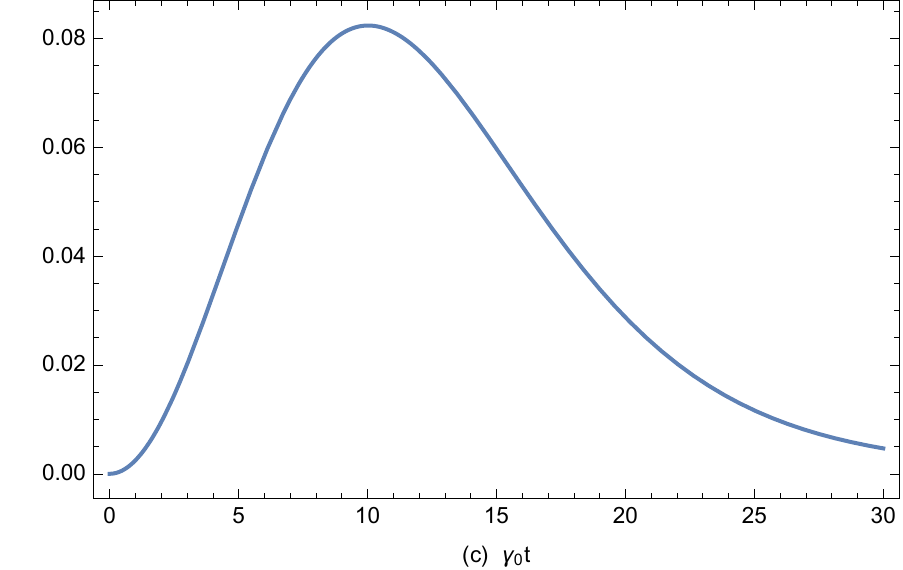}
\includegraphics[width=6cm,height=3.5cm]{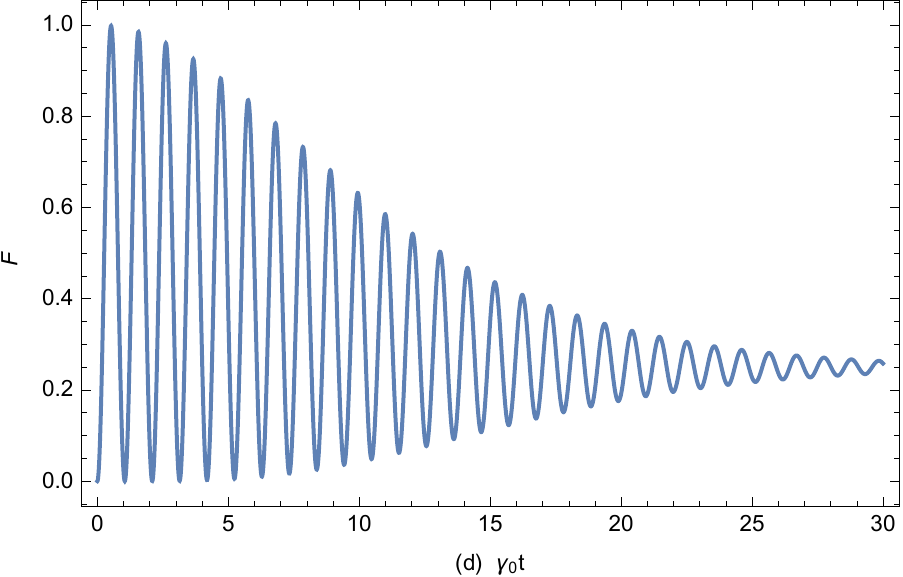}
\parbox{8cm}{\small{\bf Figure3.}
(Color online) $F_{\phi}$ as a function of the dimensionless quantity $\gamma_0t$ for the initial state $|\psi_{2}\rangle$. Here $\theta=\frac{\pi}{2}$. (a)$\lambda=5\gamma_{0}$, $\Omega=0.05\gamma_{0}$; (b)$\lambda=5\gamma_{0}$, $\Omega=3\gamma_{0}$; (c)$\lambda=0.05\gamma_{0}$, $\Omega=0.05\gamma_{0}$; (d)$\lambda=0.05\gamma_{0}$, $\Omega=3\gamma_{0}$. The inset in (b) shows the case of $\Omega=20\gamma_{0}$. }
\end{center}

\begin{center}
\includegraphics[width=6cm,height=3.5cm]{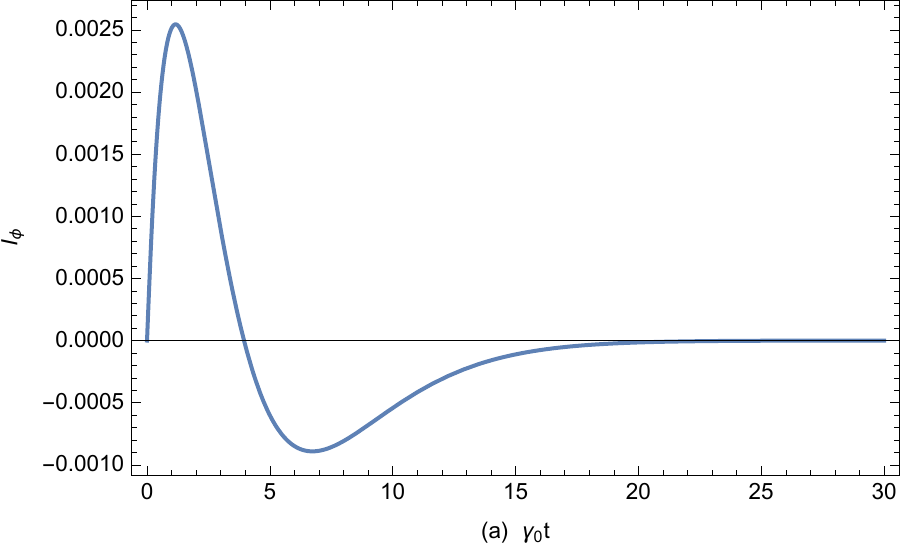}
\includegraphics[width=6cm,height=3.5cm]{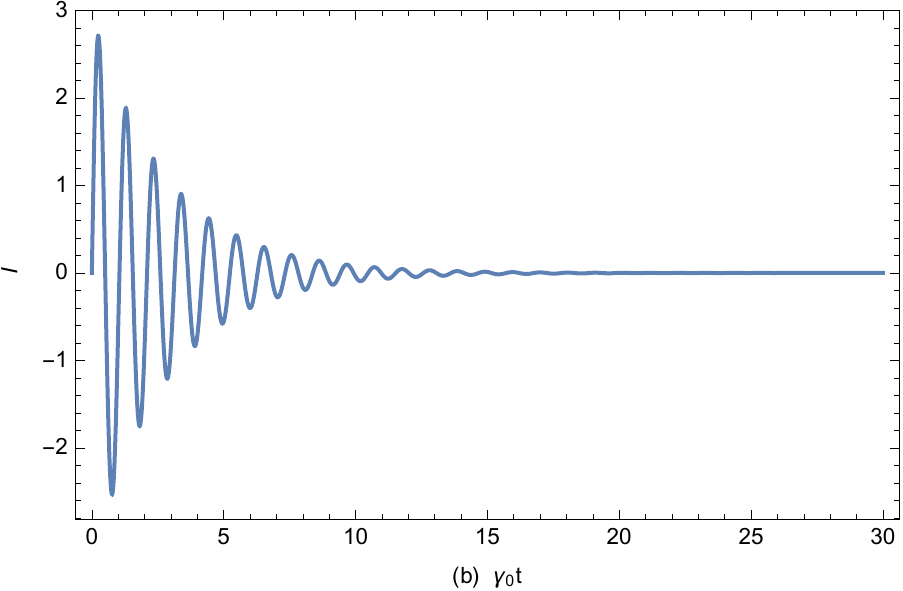}
\includegraphics[width=6cm,height=3.5cm]{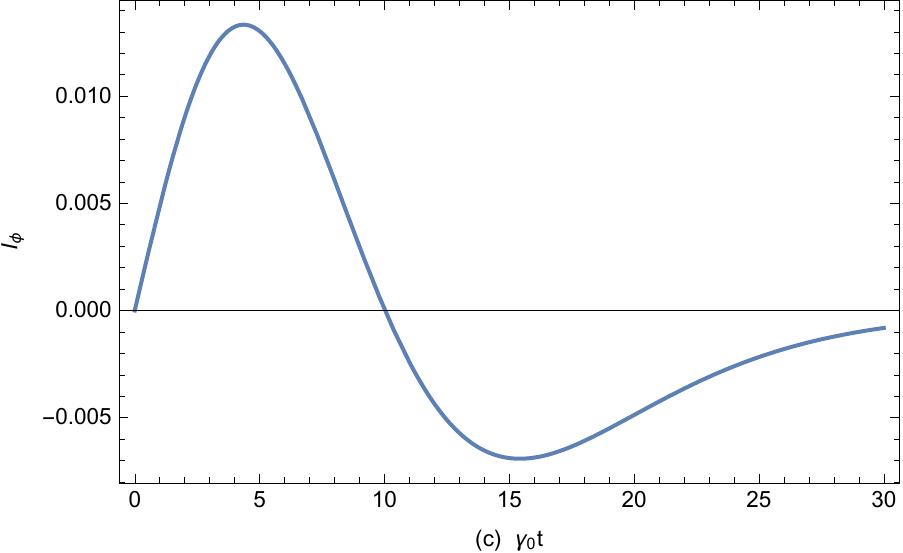}
\includegraphics[width=6cm,height=3.5cm]{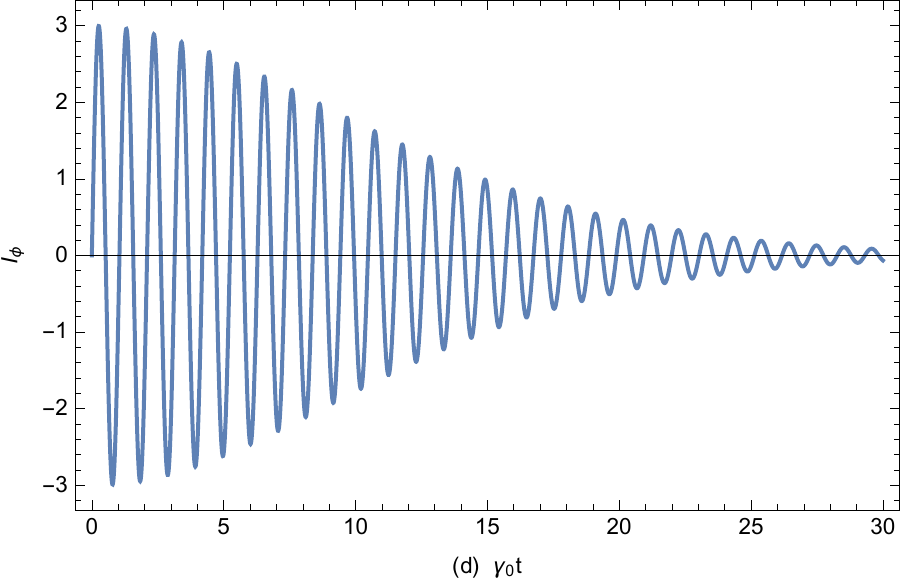}
\parbox{8cm}{\small{\bf Figure4.}
(Color online)The effect of the atom-cavity coupling $\Omega$ and cavity-reservoir coupling $\lambda$ on the QFI flow versus $\gamma_{0}t$. Here $\theta=\frac{\pi}{2}$. (a)$\lambda=5\gamma_{0}$, $\Omega=0.05\gamma_{0}$; (b)$\lambda=5\gamma_{0}$, $\Omega=3\gamma_{0}$; (c)$\lambda=0.05\gamma_{0}$, $\Omega=0.05\gamma_{0}$; (d)$\lambda=0.05\gamma_{0}$, $\Omega=3\gamma_{0}$. }
\end{center}

By analyzing the two examples, we can give a physical explanation of the above results. In the Markovian (non-Markovian) regime and with small $\Omega$, $F_{\phi}$ will obviously reduce to zero, but $F_{\phi}$ will be delayed in the non-Markovian regime, as shown Figure. 1(a)(c) and Figure. 3(a)(c). However, when $\Omega$ is very large, no matter what the Markovian or non-Markovian regime, $F_{\phi}$ will oscillate damply to the stable value, as shown Figure.1(b)(d) and Figure. 3(b)(d). Furthermore, there is difference for the second state $|\psi_{2}\rangle$ due to the memory and feedback effect of the non-Markovian environment. This difference is the change of $F_{\phi}$ from 0 to a maximum and then reduced zero. By studying $I_{\phi}$, we can understand about the change of $F_{\phi}$. The QFI flow is positive or negative, consistent with the increase or decrease of the Figure. 2 and Figure. 4. In the weak coupling regime ($\Omega=0.05\gamma_{0}$), both Markovian and non-Markovian regimes, the $I_{\phi}$ is outward, corresponding to the decay of $F_{\phi}$, as shown Figure. 2(a)(c) and Figure. 4(a)(c). However, due to the interaction with the environment, $I_{\phi}$ first changes from zero to maximum, as shown Figure. 2(a)(c). In the strong coupling regime ($\Omega=3\gamma_{0}$), $I_{\phi}$ is inward and outward as shown Figure. 2(b)(d) and Figure. 4(b)(d).

\section{Conclusion}
In conclusion, we have investigated the quantum Fisher information dynamics of the atom qubit in the dissipation cavity interacting with external environments by the TCL master equation method. We have examined two different states corresponding to two different representations: the first case is the QFI under the dressed-state basis. The second is the QFI under an arbitrary single-qubit state. The results show that the result of using different parameters is obviously biased. We demonstrated that in the weak coupling regime, the QFI about the parameter $\phi$ is monotonously decreased. When the strength of the atom-cavity coupling became stronger, the QFI will oscillate damply to the stable value of 0.25. We might consider more general parameterization than the canonical equation Eq.~(\ref{EB15}). In addition, we introduced the relationship between QFI flow and information to understand the changing trend of QFI. Thus, the QFI flow of the negative value indicates that the information flows from the system to the environment, corresponding to the QFI of decay. The QFI flow of positive value means that information flows from the environment to the system, accounting for the QFI of revival. In other words, it is important to select the appropriate parameters to improve the accuracy of parameter estimation. In the future, it will be worth to study the dynamical evolution of the QFI in the different states, and it will be also an interesting topic to explore the estimation of multiple qubit to parameter.

\begin{acknowledgments}
Project supported by the Scientific Research Project of Hunan Provincial Education Department, China (Grant No 16C0949), Hunan Provincial Innovation Foundation for Postgraduate (CX2017B177), the National Science Foundation of China (Grant No 11374096) and the Doctoral Science Foundation of Hunan Normal University, China.
\end{acknowledgments}

\end{document}